\documentclass[12pt,preprint]{aastex}
\begin{document}
\newcommand{\calu}{{\cal U}}
\newcommand{\calq}{{\cal Q}}
\newcommand{\bx}{{\rm \bf x}}
\newcommand{\bk}{{\bar{\kappa}}}
\title{Reconstruction of the One-point Distribution of
Convergence from Weak Lensing by Large-scale Structure}
\author{Tong-Jie Zhang}
\affil{Department of Astronomy, Beijing Normal University, Beijing
100875, P.R.China; tjzhang@bnu.edu.cn; and Canadian Institute for
Theoretical Astrophysics, University of Toronto, M5S 3H8, Canada;
tzhang@cita.utoronto.ca; and National Astronomical Observatories,
Chinese Academy of Sciences, Beijing 100012, China}
\author{Ue-Li Pen}
\affil{Canadian Institute for Theoretical Astrophysics, University
of Toronto, M5S 3H8, Canada; pen@cita.utoronto.ca;
and National Astronomical Observatories,
Chinese Academy of Sciences, Beijing 100012, China}

\begin{abstract}

Weak lensing measurements are starting to provide statistical maps
of the distribution of matter in the universe that are
increasingly precise and complementary to cosmic microwave
background maps. The probability distribution (PDF) provides a
powerful tool to test non-Gaussian features in the convergence
field and to discriminate the different cosmological models. In
this letter, we present a new PDF space Wiener filter approach to
reconstruct the probability density function of the convergence
from the noisy convergence field. We find that for parameters
comparable to the CFHT legacy survey, the averaged PDF of the
convergence in a 3 degree field can be reconstructed with an
uncertainty of about 10$\%$, even though the pointwise PDF is
noise dominated.

\end{abstract}


\section{Introduction}

Mapping the mass distribution of matter in the Universe has been a
major challenge and focus of modern observational cosmology. There
are very few direct ways to weigh the universe. The only direct
procedure to weigh the matter in the universe is by using the
deflection of light by gravity. Weak gravitational lensing
provides a direct statistical measure of the dark matter
distribution regardless of the nature and dynamics of both the
dark and luminous matter intervening between the distant sources
and observer. Weak lensing by large-scale structure can lead to
the shear and magnification of the images of distant faint
galaxies. While this effect is very small, a large statistical
sample can provide a precise measurement of averaged quantities.

Based on the theoretical work done by \cite{1967ApJ...147...61G},
\cite{1991MNRAS.251..600B}, \cite{1991ApJ...380....1M} and
\cite{1992ApJ...388..272K} performed the first calculation of weak
lensing by large-scale structure, the result of which showed the
expected distortion amplitude of weak lensing effect is at a level
of roughly a few percent in adiabatic cold dark matter models.
\cite{1992ApJ...388..272K} also made early predictions for the
power spectrum of the shear and convergence using linear
perturbation theory. Due to the weakness of the effect, all
detections have been statistical in nature, primarily in regimes
where the signal-to-noise is less than unity.  Fortunately,
several groups have been able to measure this weak gravitational
lensing effect
\citep{2000MNRAS.318..625B,2002ApJ...572L.131R,2002ApJ...572...55H,
2002A&A...393..369V,2003AJ....125.1014J,2003MNRAS.341..100B,
2003ApJ...597...98H} in recent years.

In the standard model of cosmology, fluctuations start off small,
symmetric and Gaussian.  Even in some non-Gaussian models like
topological defects, initial fluctuations are still symmetric:
positive and negative fluctuations occur with equal probability
\citep{1994PhRvD..49..692P}.  As fluctuations grow by
gravitational instability, this symmetry can no longer be
maintained - over densities can be arbitrarily large, while under
dense regions can never have less than zero mass.  This leads to a
skewness in the distribution of matter fluctuations.
\cite{2003ApJ...592..664P} have measured the first detection of
dark matter skewness from the VIRMOS-DESCART survey. They find the
lensing skewness can be detected to be  for a compensated Gaussian
filter on scales of 5.37. Using the skewness of the convergence
for simulated weak lensing, \cite{2003ApJ...598..818Z} present the
first optimal analysis for the
Canada-France-Hawaii-Telescope(hereafter CFHT) Legacy Survey in
the presence of noise due to the randomly oriented intrinsic
ellipticity of source galaxies.
%\cite{2003ApJ...598..818Z} present the first extended comparison
%for different kinds of window functions to isolate the filter
%that is optimal for distinguishing cosmological models.
They show that a compensated Gaussian filter on a scale of 2.5 arc
minutes can optimizes cosmological constraint with $\Delta
\Omega_m/\Omega_m\sim 10\%$, which is significantly better than
other window function that have been considered in the literature.
While skewness has already been measured at very high statistical
significance \citep{2003ApJ...592..664P}, the measurement has not
resulted in a strong constraint on the total matter density
$\Omega_m$. The data has so far been limited by sample variance
and analysis techniques. Currently ongoing surveys, such as CFHT
Legacy Survey, will provide more than an order of magnitude
improvement in the statistics.

Besides the skewness of the convergence, the probability density
distribution (hereafter PDF) provides a powerful tool to test
non-Gaussian features of the convergence field and to discriminate the
different cosmological models \citep{2000ApJ...530..547J}.  In real
data of weak lensing, the noise due to the intrinsic ellipticity of
background galaxies overwhelms the PDF of convergence. So far there is
no good method to extract the underlying PDF of the convergence from
the observed noisy version. In this letter, we present a new Wiener
reconstruction approach to reconstruct the PDF distribution of
noise-free convergence field from the convergence field added with a
Gaussian noise field.

The outline of the paper is as follows. In \S 2, we introduce the
strategy of map construction of weak lensing from simulations. In
\S 3, we describe the PDF Wiener method of reconstructing the
distribution of the convergence, while we present the results and
summarize our conclusions in \S 4 and \S 5 respectively.

\section{ Map Construction of Weak Lensing by Simulations}

A full non-Gaussian statistical description of the distribution of
matter is very complex.  It would require the joint n-point
distribution function of matter, which is difficult to describe or
compute. A drastic simplification occurs if we only consider the
one point distribution of the density field.  In gravitational
lensing, one only measures the projected dark matter density.  We
then smooth the field with some appropriate filter, and consider
the one point PDF of the smoothed density field.  Of particular
interest is the deviation of this one point PDF from Gaussianity.

A separate question is the degree of information loss implied by
this assumption.  This can be quantified in the context of errors
on cosmological parameters in a parameterized cosmological model.
In a future paper, we will present the accuracy with respect to a
set of parameters. In this letter, we will simply address the
challenge of reconstructing the underlying PDF from a noisy
measurement.  In a controlled simulation, we compare the
reconstructed PDF to the underlying noise-free PDF.

\subsection{N-Body Simulation}

We ran an N-body simulation with a cosmological model of WMAP to
generate convergence maps. The power spectra for given parameters
were generated using CMBFAST \citep{1996ApJ...469..437S} and these
tabulated functions were used to generate initial conditions.  The
power spectra were normalized to be consistent with the earlier
two point analysis from this data set \citep{2002A&A...393..369V}.
We ran the simulations using a parallel, Particle-mesh N-body code
PMFAST \citep{2004astro.ph..2443M} at $1856^3$ mesh resolution
using $928^3$ particles on an 4 node quad processor Itanium
Beowulf cluster at CITA. This publicly available N-body code
features very fast execution and negligible memory overhead: only
the positions and velocity of particles are stored, requiring six
numbers.  One more integer is used for every particle to store
linked lists.  All other arrays are small and recycled.  Parallel
execution efficiency is achieved by decomposing the gravitational
force into a long range and short range component. The long range
force is solved on a coarse grid, which requires very little
computing or communications resources.  The fine grid is only
stored in patches, and is used to compute the short range force.
The highly optimized vendor IPP FFT library is used.

Output times were determined by the appropriate
tiling of the light cone volume with joined co-moving boxes from
$z \approx 3$ to $z=0$. We output periodic surface density maps at
$1856^2$ resolution along the 3 independent directions of the cube
at each output interval. These maps represent the raw output for
the run and are used to generate convergence maps in the thin lens
and Born approximations by stacking the images with the
appropriate weights through the comoving volume contained in the
past light cone.

The simulation started at an initial redshift $z_i=80$, and ran
for 285 steps in equal expansion factor ratios with box size
$L=100 h^{-1}$ Mpc comoving.  We adopted a Hubble constant
$h=0.71$, and a flat
cosmological model with $\Omega_m+\Omega_{\Lambda}=1$ was used.  We
used the WMAP parameters of $\Omega_m=0.27$ \citep{2003ApJS..148....1B}.
The power spectrum normalization $\sigma_8$ was chosen as 0.84.
%, and primordial power law index $n=1$.

\subsection{Construction of Simulated Convergence Maps}\

The convergence $\kappa$ is the projection of the matter
over-density $\delta$ along the line of sight $\theta$ weighted by
the lensing geometry and source galaxy distribution. It can be
expressed as
\begin{equation}
\kappa(\theta,\chi_s)=\int_0^{\chi_s} W(\chi) \delta(\chi,r(\chi)
\theta))d\chi,
\end{equation}
where, $\chi$
is the comoving distance in unit of $c/H_0$, and $H_0=100\  h
\ {\rm km/s/Mpc}$. The weight
function $W(\chi)$ is\
\begin{equation}
W(\chi)=\frac{3}{2}\Omega_{m}g(\chi)(1+z)
\end{equation}
determined by the source galaxy distribution function $n(z)$ and
the lensing geometry.
\begin{equation}
g(\chi)=r(\chi) \int_{\chi}^{\infty} d\chi' n(\chi')\frac{r(\chi'
-\chi)}{r(\chi')}.
\label{eqn:gx}
\end{equation}
$r(\chi)$ is the radial coordinate. $r(\chi)=\sinh(\chi)$ for
open, $r(\chi)=\chi$ for flat and $r(\chi)=\sin(\chi)$ for closed
geometry of Universe.  In our computations, we adopt a flat geometry.
$n(z)=n(\chi)d\chi/dz$ is normalized such
that $\int_0^{ \infty} n(z)dz=1$. For the CFHT Legacy Survey, we
adopt
$n(z)=\frac{\beta}{z_0\Gamma(\frac{1+\alpha}{\beta})}(\frac{z}{z_0})^
\alpha\exp(-(\frac{z}{z_0})^\beta)$ with $\alpha=2$ and
$\beta=1.2$ and the source redshift parameter $z_0=0.44$, which
peaks at $z_p=1.58z_0$, respectively. The mean redshift is
$\bar{z}=2.1z_0$ and the median redshift is
$z_h=1.9z_0$\citep{2002A&A...393..369V}. The source redshift
distribution $n(z)$ adopted here is the same as that for VIRMOS.

\begin{figure}
\epsscale{0.9}\plotone{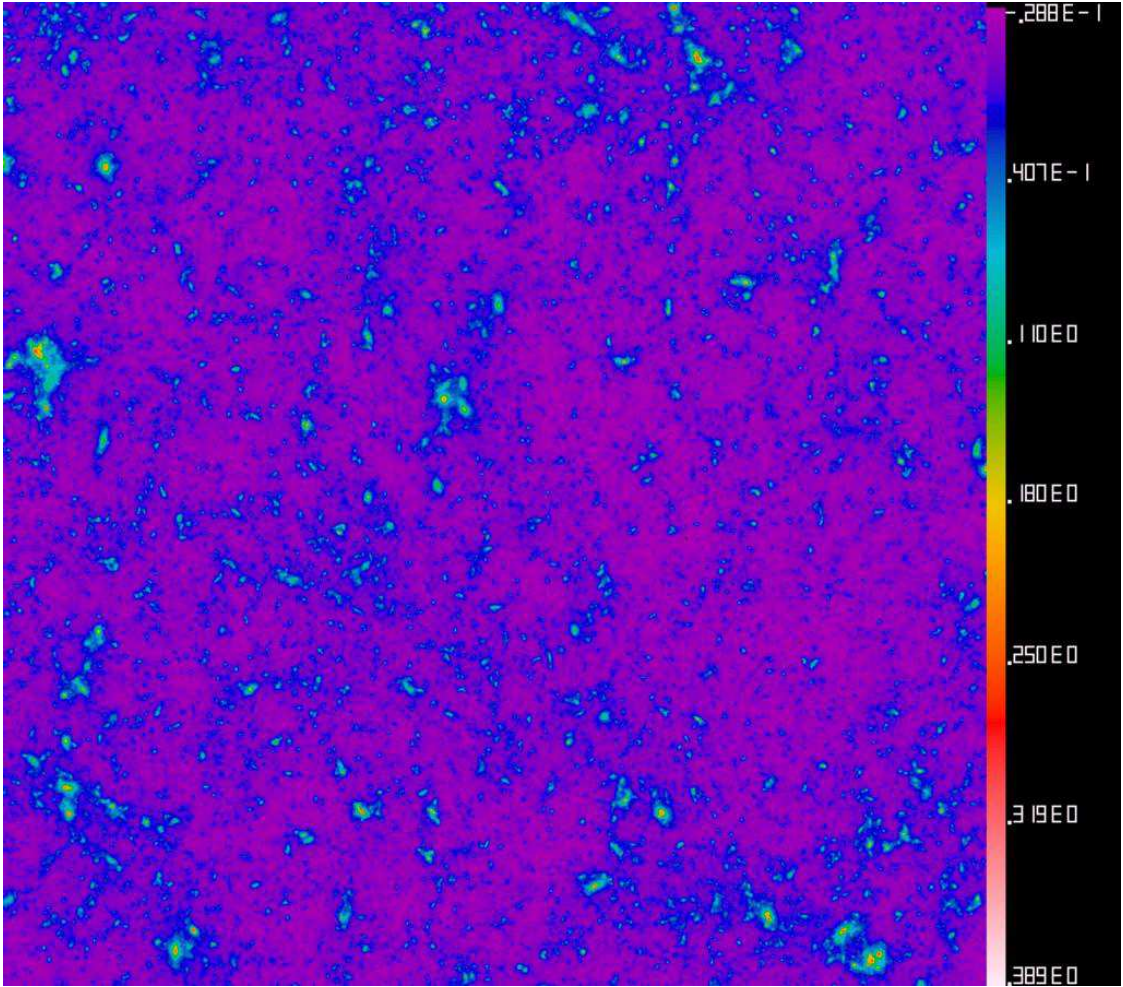}\caption{A noise-free $\kappa$ map
in the N-body simulation of a WMAP cosmology with a width of 3
degrees and $2048^2$ pixels, and the scale is in units of
$\kappa$.} \label{fig1}
\end{figure}

During each simulation we store 2D projections of $\delta$ through
the 3D box at every light crossing time through the box along all
x, y and z directions. Our 2D surface density sectional maps are
stored on $1856^2$ grids. After the simulation, we stack sectional
maps separated by a width of the simulation box, randomly choosing
the center of each section and randomly rotating and flipping each
section.  The periodic boundary condition guarantees that there is
no discontinuities between any two adjacent boxes. We then add
these sections with the weights given by $W(z)$ onto a map of
constant angular size, which is generally determined by the
maximum projection redshift. To minimize the repetition of the
same structures in the projection, we alternatively choose the
sectional maps of x, y, z directions during the stacking. Using
different random seeds for the alignments and rotations, we make
$40$ maps for each cosmological model. Since the galaxy
distribution peaks at $z\sim 1$, the peak contribution of lensing
comes from $z\sim 0.5$ due to the lensing geometry term. Thus the
maximum projection redshift $z\sim 2$ is sufficient for the
lensing analysis. So we project the $\Omega_m=0.27$ simulations to
$z=2$ and obtain $40$ maps each with angular width
$\theta_\kappa=3$ degrees.  At the very highest redshift bins, the
field of view is larger than the simulation volume.  In those slices
we used periodic boundary conditions to fill in the field.  Their
lensing contributions are negligible.
Fig.(\ref{fig1}) shows a noise-free
$\kappa$ map in the N-body simulation of a WMAP cosmology with a
width of 3 degrees and $2048^2$ pixels, and the scale is in units
of $\kappa$.

We then simulate the CFHT Legacy Survey by adding noise to these
clean maps. The noise $\kappa$ maps have a pixel-pixel variance
$\sigma_N^2=\langle e^2\rangle/2/\langle N_{\rm
pixel}\rangle=0.74^2$, which is the noise variance in each pixel
before wiener filter. Here $\langle e^2\rangle=0.47^2$ is the total
noise estimated in the VIRMOS-DESCART survey and here we take it as
what would be expected by the CFHT Legacy Survey. It includes the
dispersion of the galaxy intrinsic ellipticity, PSF correction noise
and photon shot noise. $\langle N_{\rm pixel} \rangle$ is the mean
number of galaxies in each pixel. For VIRMOS, the number density of
observed galaxies $n_g\simeq 26/{\rm arc min}^2$, then $\langle
N_{\rm pixel} \rangle=n_g[(\theta_\kappa/1^{'})/N]^2$, where
$N=2048$ is the number of grids by which we store 2D maps and the
field of view $\theta_\kappa$ is in units of arc min. The factor of
2 arises from the fact that the shear field has two degrees of
freedom ($\gamma_1,\gamma_2$), where the definition of $\langle
e^2\rangle$ sums over both. We use this as our best guess for the
CFHT Legacy Survey noise. We smooth both of the noise-free and noisy
convergence fields by a Wiener filter $F_W$, so $\kappa_W(x)=\int
\kappa(x') F_W(x-x')$.  The Fourier transform of the Wiener filter
is given by
\begin{equation}
F_W(k)=\frac{P^S(k)}{P^S(k)+P^N(k)}
\end{equation}
where $P^S(k)$ is the power spectrum of the $\kappa$ field and
$P^N(k)$ is the power spectrum of the noise.
The broad Wiener filter smoothes the image, and reduces the variance
of the noise per pixel. This can be revealed by the dotted line
in Fig.(\ref{fig4}), where the variance of pure Gaussian noise field
smoothed with the Wiener filter is $\sigma_{Nf}=0.0084$.
The Wiener filter has an effective area of
$(\sigma_N/\sigma_{Nf})^2$ pixels. Of course, the pixels
are highly correlated now.

The real $\kappa$ field is quite non-Gaussian, and a Wiener filter
might not be the best.  An alternative decomposition based on wavelets
is discussed by \cite{pen1999}.

\begin{figure}
\epsscale{0.9} \plotone{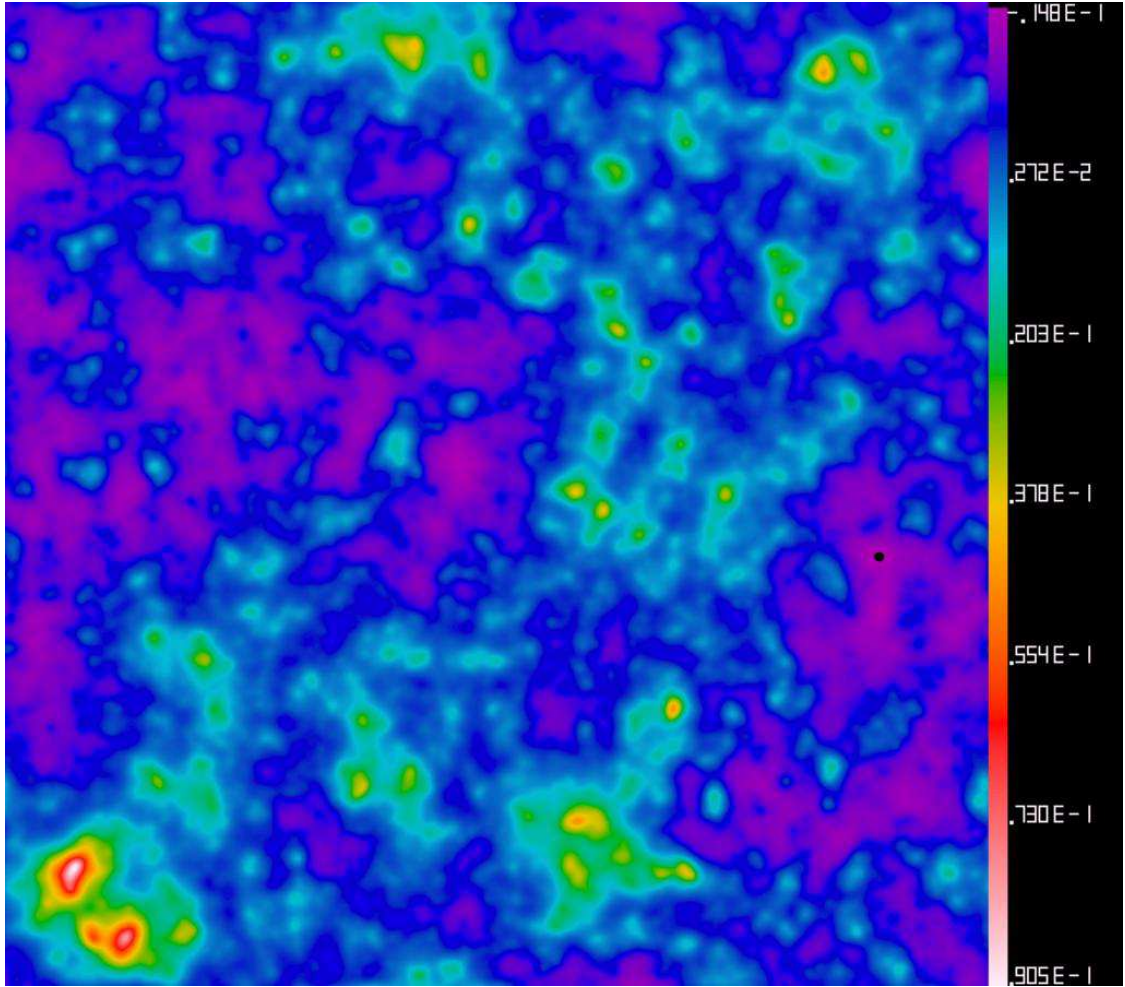} \caption{The same as
Fig.\ref{fig1}, but for the noise-free $\kappa$ map smoothed by a
Wiener filter.} \label{fig2}
\end{figure}

\begin{figure}
\epsscale{0.9} \plotone{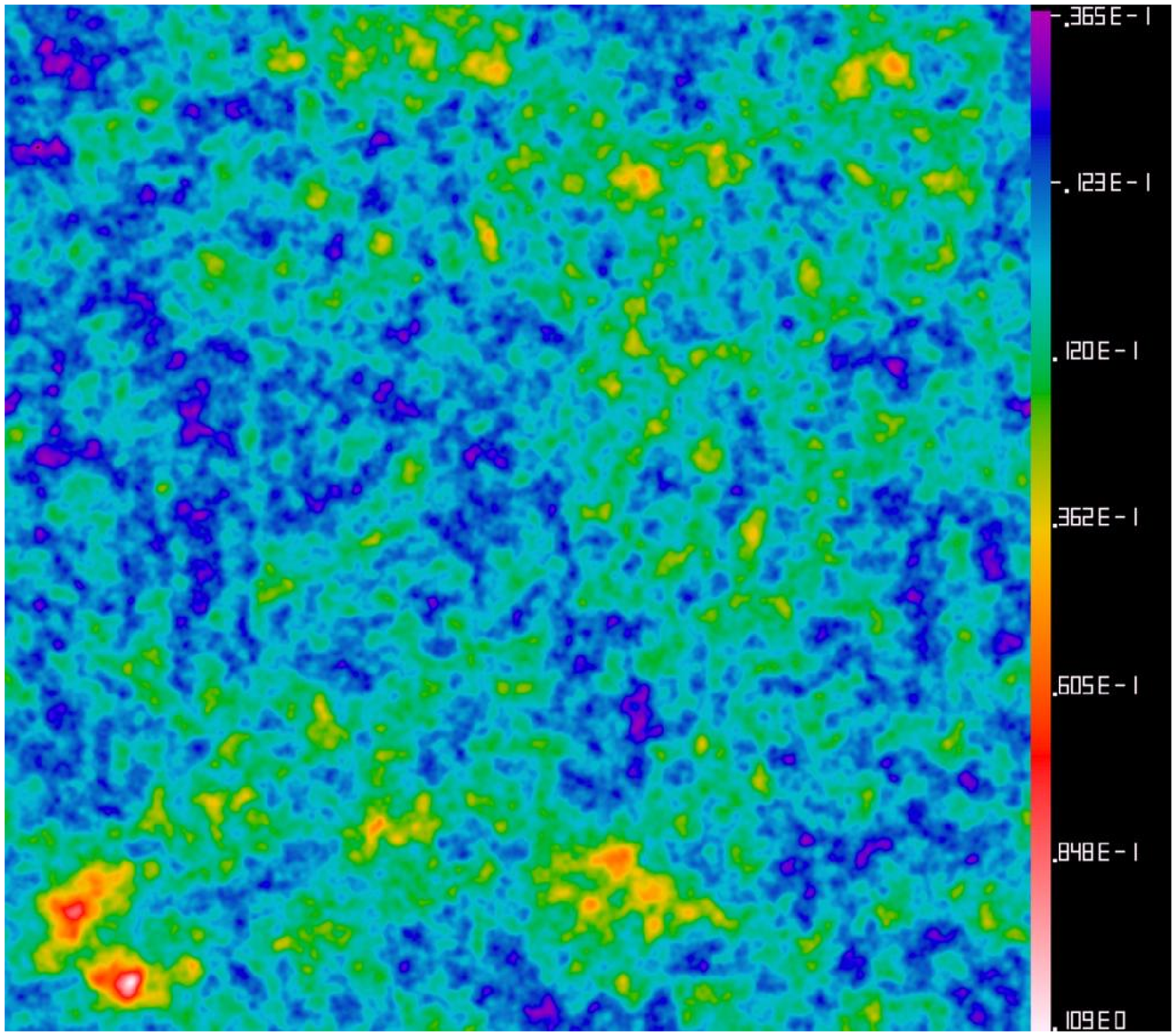} \caption{The same as
Fig.\ref{fig2}, but for noisy $\kappa$ map by adding a simulated
Gaussian noise field smoothed with the same Wiener filter.}
\label{fig3}
\end{figure}

Fig.(\ref{fig2}) and (\ref{fig3}) show the the noise-free and noisy
$\kappa$ maps by adding simulated Gaussian noise field smoothed by
Wiener filter respectively.  It is apparent that even the minimum
noise Wiener maps are very noisy, and where noise dominates salient
features in the maps. The $\kappa$ field is equivalent to the E-mode
of the shear field ($\gamma$), and a Wiener filter allows a direct
mapping of the shear to filtered $\kappa$ field
\citep{2003NewA....8..581P}. It is well known that shear and kappa
have equivalent noise properties \citep{1998ApJ...506...64S}. This
can be proven as follows: white and independent noise in $\gamma_1$
and $\gamma_2$, when Fourier transformed, is still white. The
decomposition into E and B modes, i.e. the component of the shear in
Fourier space that is parallel and perpendicular to the wave vector,
is just a rotation of two uncorrelated noise variables.  This gives
two new uncorrelated noise variables of equal variance.  The E mode
and B mode will both have white noise again, which is statistically
identical to the noise in $\gamma_1$ and $\gamma_2$ (differing by
only a rotation). The E-mode is identical to kappa, except for one
constant mode. Therefore kappa and shear can have their noise
treated identically. For real surveys, the naive Wiener filter costs
as the number of galaxies to the third power, and could be
computationally very expensive, but fast $O(N)$ algorithms have
already been demonstrated \citep{2003MNRAS.346..619P}.

\section{Wiener Reconstructing of the Convergence PDF}

We apply Wiener filtering in two completely different contexts.  In
the previous sections, the Wiener filter was used to generate a
minimum noise variance map.  The next step is the tabulation of the
PDF from a 9 square degree simulated survey field.  This PDF is
necessarily a noisy realization of the ensemble expectation value.

In addition to the sample variance noise, the observations themselves
have noise.  This is noise is added pointwise to the intrinsic lensing
surface density.  The PDF of the sum of signal plus noise is the
convolution of the two PDF's.  We model the noise PDF as a Gaussian
with zero mean and specified standard deviation.  We further assume
that the noise standard deviation is known.

In the ensemble average, one could in principle recover the lensing
PDF from the noisy PDF by a deconvolution.  In practice, of course,
this is very difficult because the small scale noise is amplified by
the deconvolution to the extent that it completely overwhelms the
signal.  \cite{2002ApJ...578L..27L} showed a heuristic scheme for noisy
Sunyaev-Zeldovich measurements, which regularizes this deconvolution
by applying a Wiener filter on the PDF of peaks.  Lee extracted the ``power
spectrum'' of the signal and noise in the PDF using Fourier
transforms.  Fourier transforms are good descriptions for stationary
processes, but clearly a PDF is not a stationary process as a function
of $\kappa$.

Stationarity is not a requirement for Wiener filtering.  One can
specify the full covariance matrix.  However, one must be careful to
formulate the question.  What is meant by covariance?  What is one
approximating by a Gaussian process?

We first consider the literal interpretation.  We define the ensemble
average PDF of the 40 maps as the underlying PDF.  The actual PDF of
each map is a noisy realization.  The difference of each map PDF from
the ensemble average is the noise free signal variance.

Next, we consider a single map PDF with noise added.  If the process
of adding noise were an exact convolution, this would be a reversible
process for a non-zero kernel, in particular for a Gaussian.  To
assess the actual contribution of noise, we take the noise-free PDF
$P^S$, and convolve by the ideal noise kernel.  Then we take the noisy
map, and measure its actual noisy PDF $P^{S+N}$.  The difference of
the noisy map's PDF from the ideal convolved PDF is a measure of the
noise's contribution on the irreversibility of the convolution.  We
thus define the noise covariance matrix as the covariance of the
difference between the ideal convolved PDF and the noisy PDF of each
map.  The Wiener filtered deconvolution can then be readily applied to
the observed noisy PDF, as we will demonstrate below.

One might wonder how circular this procedure really is.  Our ultimate
goal is to find a minimum noise measurement of the ensemble PDF from a
noisy measurement of a single field PDF.  In order to implement this,
we had assumed prior knowledge of an ensemble of fields, as well as
knowledge of the ensemble average $\kappa$ PDF and its covariance!  At
first sight this seems to imply much more prior knowledge than we set
out to solve for.  In practice, the problem is not as serious as it
might sound.  It needs to be shown that the actual assumed PDF for the
signal is not important, or that it can be derived from the observed
data alone.

We thus test not only the literal interpretation of a Wiener
reconstructed PDF, but also of a more liberal version, we have assumed
the signal PDF to be Gaussian.  For a Gaussian, only the variance needs
to be known or measured from the data, which can already be done to
good accuracy from existing lensing data.  The quality of the
reconstruction will necessarily be worse, but we will show from actual
simulations that good results can nevertheless be achieved.

The PDFs of pure $\kappa$ field, pure noise and $\kappa$ field added
by noise are $P^S(\kappa)$, $P^N(\kappa)$ and $P^{S+N}(\kappa)$
respectively.  The number of maps is $N=40$, i.e. $n=1,2,\cdots N$.
Normally, PDFs are normalized to be unit, i.e.
$\int^{+\infty}_{-\infty}P(\kappa)d\kappa=1$.  From here on we only
consider binned PDFs, choosing $M$ equally spaced bins separated by
$\Delta\kappa$.  Our notation becomes $\kappa_i =\kappa_-+i
\Delta\kappa$, $P_i \equiv P(\kappa_i) \Delta \kappa$, and each PDF
can also carry an appropriate superscript. The number of bins for
all PDFs is $M=19$, i.e. $i,j=1,2, \cdots M$.

In the form of matrix, $\mathbf{C}^S$ is the signal covariance and
$\mathbf{C}^N$ is a
noise covariance. The signal covariance is expressed as
\begin{equation}
C^S_{ij}=<[P^S_i-\bar{P}^S_i][P^S_j-\bar{P}^S_j]>,
\label{eqn:cs}
\end{equation}
where $<>$ is the average over 40 maps, and $\bar{P}^S(\kappa)$ is
the average value of the $\kappa$ field PDF for 40 maps.
The noise PDF, which will also serve as the noise convolution kernel
$g^N(\kappa-\kappa')$,  is a Gaussian distribution
\begin{equation}
g^N(\kappa_i-\kappa_j)=\frac{1}{\surd{2\pi}\sigma_N}
\exp\left[-\frac{(\kappa_i-\kappa_j)
^2}{2\sigma^2_N}\right],
\end{equation}
where $\sigma_N$ is the standard deviation of the noise after the
wiener filter. We can also think of this kernel as a noise matrix
$\mathbf{N}$
\begin{equation}
N_{ij}=g^N(\kappa_i-\kappa_j)\Delta \kappa.
\end{equation}
For a noisy $\kappa$ field, the noise deviation
of the PDF relative to an invertible convolution can be written as for each map
\begin{equation}
\Delta P^N_i\equiv P^{S+N}_i-N_{ij}P^S_j,
\end{equation}
where $P^{S+N}(\kappa_i)$ is the PDF of $\kappa$ field with added
noise. Thus we have the noise covariance matrix
\begin{equation}
C^N_{ij}=<[\Delta P^N_i-\Delta\bar{P}^N_i][\Delta P^N_j-\Delta\bar{P}^N_j]>.
\end{equation}

Using the convolution theorem, we have
\begin{equation}
P^{S+N}=\int^{+\infty}_{-\infty}P^S(\kappa')g^N(\kappa-\kappa')d\kappa,
\end{equation}
\begin{equation}
P^{S+N}_i=\sum_{j=1}^M N_{ij}\cdot P^S_j,
\end{equation}
which reduces to in the form of matrix
\begin{equation}
\mathbf{\tilde{P}}^{S+N}=\mathbf{N}\mathbf{\tilde{P}}^S,
\end{equation}
so we have
\begin{equation}
\mathbf{\tilde{P}}^S=\mathbf{N}^{-1}\mathbf{\tilde{P}}^{S+N}.
\end{equation}
The Wiener filter is defined as,
\begin{equation}
\mathbf{W}=\mathbf{C}^S(\mathbf{C}^S+\mathbf{\tilde{C}}^N)^{-1},
\end{equation}
where
$\mathbf{\tilde{C}}^N=\mathbf{N}^{-1}\mathbf{C}^N\mathbf{N}^{-1}$.

%the PDF of reconstructed signal can be obtained
%\begin{equation}
%\mathbf{\tilde{P}}^S_W=\mathbf{W}\mathbf{\tilde{P}}^S=\mathbf{W}
%\mathbf{N}^{-1}\mathbf{\tilde{P}}^{S+N}=\mathbf{C}^S
%\mathbf{N}[\mathbf{N}\mathbf{C}^S\mathbf{N}+\mathbf{C}^N]^{-1}
%\mathbf{\tilde{P}}^{S+N}.
%\end{equation}

All covariances are specified as deviations from a known underlying
model.  Given a the PDF of a noisy map, we then must first subtract
the ensemble average noisy PDF, and apply the Wiener reconstruction to
difference.  After a applying a Wiener filtered deconvolution, we add
back the noise free ensemble average PDF.
The figure of merit is how well the reconstruction finds
the noise-free PDF of the original map (not the ensemble average).

The reconstruction is thus decomposed into two parts: an underlying
mean noisy PDF has a known pre-convolved (i.e. deconvolved) PDF.  The
difference between the map realization and this ensemble average is
then deconvolved with a Wiener filter.

Expressed as equations, the deviation of reconstructed
$\mathbf{\tilde{P}}^S_W$ is
\begin{equation}
\Delta\mathbf{\tilde{P}}^S_W=\mathbf{C}^S
\mathbf{N}[\mathbf{N}\mathbf{C}^S\mathbf{N}+\mathbf{C}^N]^{-1}
\Delta\mathbf{\tilde{P}}^{S+N}
\end{equation}
where
\begin{equation}
\Delta\mathbf{\tilde{P}}^{S+N}=\mathbf{\tilde{P}}^{S+N}-
\mathbf{\bar{\tilde{P}}}^{S+N};\ \ \mathbf{\bar{\tilde{P}}}^{S+N}
=\mathbf{N}\mathbf{\bar{\tilde{P}}}^S.
\label{eqn:dp}
\end{equation}
Thus we can obtain the reconstructed $\mathbf{\tilde{P}}^S_W$
\begin{equation}
\mathbf{\tilde{P}}^S_W=\Delta\mathbf{\tilde{P}}^S_W+\mathbf{\bar{\tilde{P}}}^S,
\end{equation}
%\begin{equation}
%\bar{\tilde{P}^{S+N}(\kappa_j)}=N_{ij}\bar{P^S(\kappa_j)}
%\end{equation}
and its error
\begin{equation}
\mathbf{\sigma}^2_{\mathbf{\tilde{P}}^S_W}=<(\mathbf{\tilde{P}}
^S_W-\mathbf{\tilde{P}}^S)^2>,
\end{equation}
where $<>$ is the mean over 40 maps, and $\mathbf{\tilde{P}}^S$ is the PDF
of the $\kappa$ field.

A key assumption was that we actually knew the ensemble average PDF.
The difference between the observed PDF and the ensemble average is a
smaller amplitude function, and can be deconvolved using an
appropriate Wiener filter.  In practice, one might not know the
underlying PDF, and subtract the wrong function.  To test this
possibility, we also used a Gaussian mean distribution for
$\mathbf{\bar{\tilde{P}}}^S$ in equation (\ref{eqn:dp}).  In
the reconstructed PDF we again used the exact deconvolution of the
Gaussian (which is of course also a Gaussian)
$\mathbf{\bar{\tilde{P}}}^S$. In this case, we are applying the exact
deconvolution to a lesser component of the observed PDF, and test how
important it is to know the true underlying ensemble average PDF. One
could in principle also use the signal covariance field of a Gaussian
field in Equation (\ref{eqn:cs}).  It does not appear as crucial a
qualitative difference, and was not tested.

%For comparison, we also replace $\bar{P}^S(\kappa_j)$ by a
%Gaussian distribution with the standard deviation
%$\mathbf{\bar{\sigma}}_\kappa$ of 40 $\kappa$ fields. We do not
%obtain the pdf $\bar{P}^S(\kappa_j)$ of real pure $\kappa$ field,
%but we can measure the standard deviation
%$\mathbf{\bar{\sigma}}_\kappa$.

\section{The reconstructed probability density function (PDF)
and the cumulative distribution function (CDF)}

\begin{figure}
\epsscale{1.} \plotone{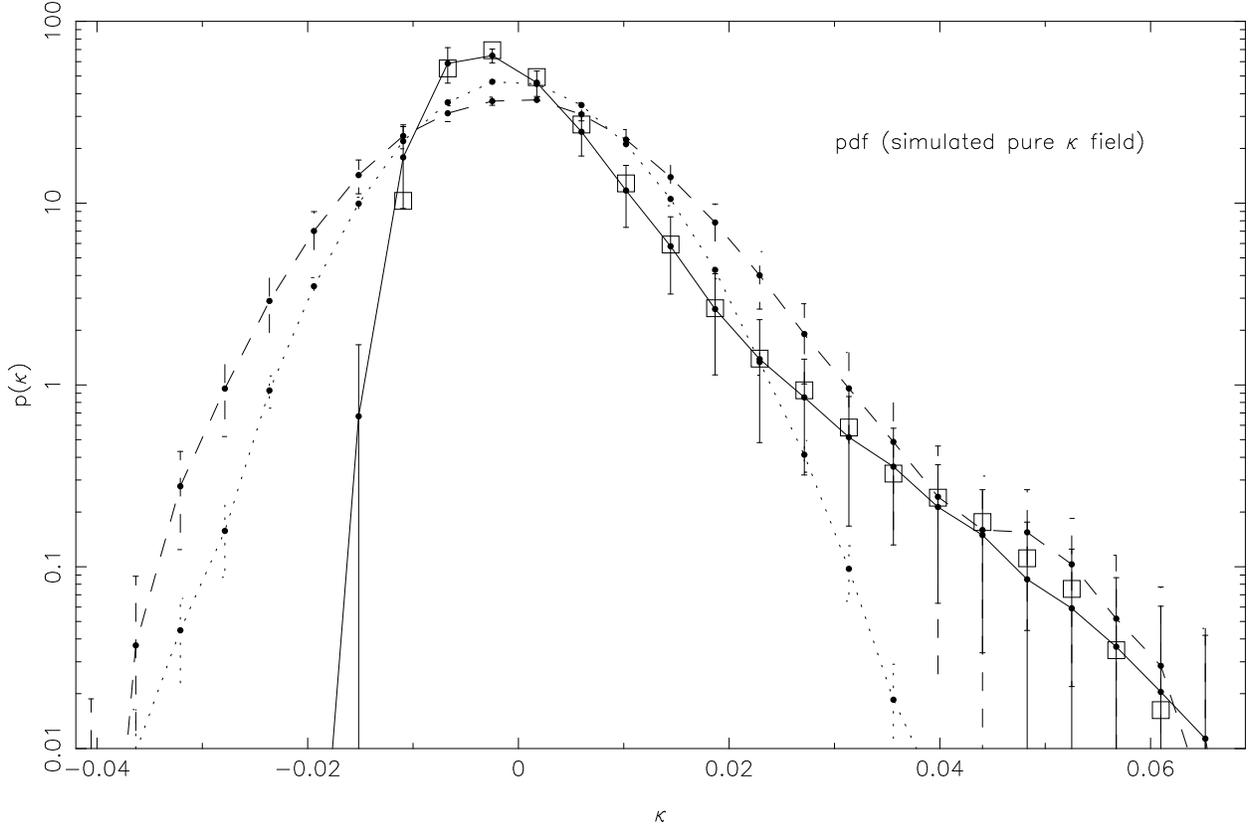} \caption{The probability
distribution functions (PDF) of the noise-free $\kappa$ field
(solid line), noisy $\kappa$ field (dashed line) and pure Gaussian
noise field (dotted line) respectively. The corresponding error
bars are obtained from averaging over 40 respective maps. The
empty square dots represent the reconstructed PDF of $\kappa$
field using simulated pure $\kappa$ field as a initial signal
field.} \label{fig4}
\end{figure}

\begin{figure}
\epsscale{1.} \plotone{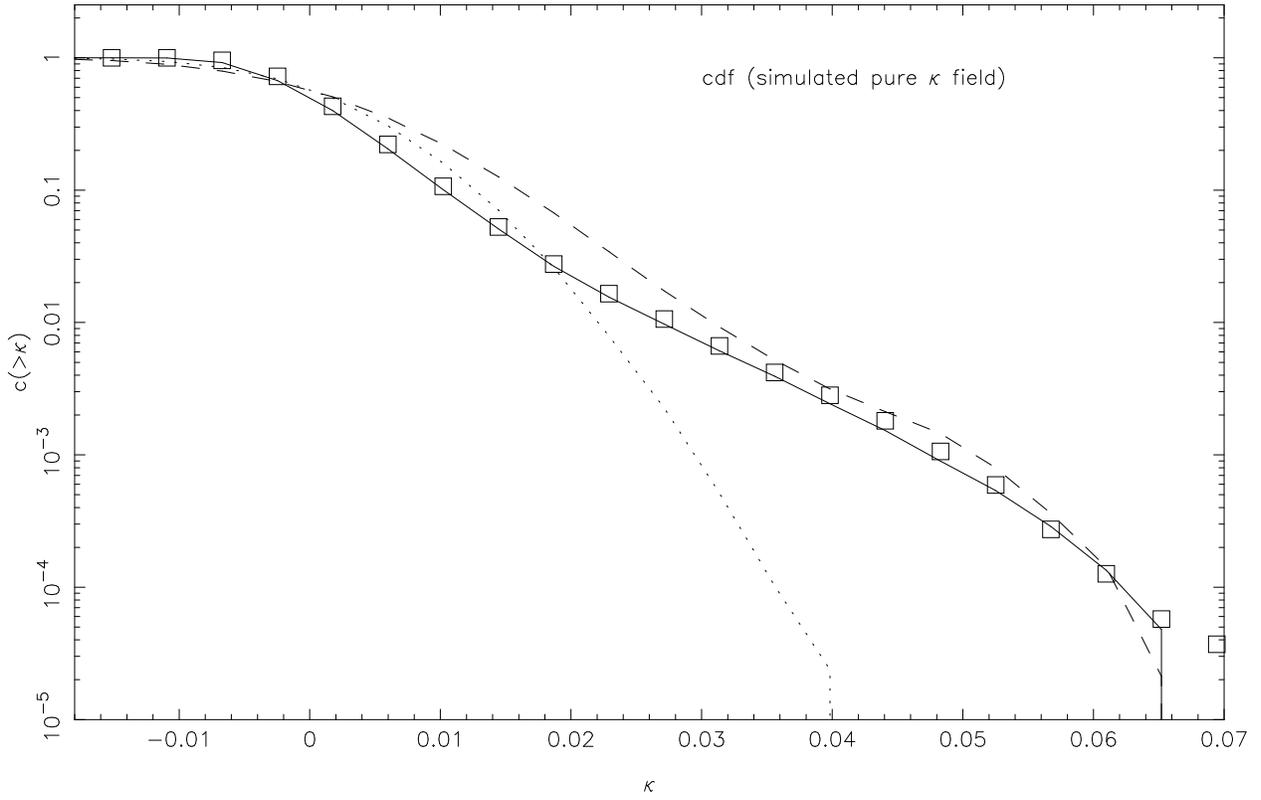} \caption{The same as
Fig.\ref{fig4}, but for the corresponding cumulative distribution
functions (CDF).} \label{fig5}
\end{figure}

The PDF of a convergence map is defined as
\begin{equation}
P(\kappa)=\frac{n(\Delta\kappa)}{N^2\Delta\kappa},
\end{equation}
where $N$ is the number of grids by which 2D maps are stored and
$n(\kappa)$ is the number of the within unit interval
$\Delta\kappa$ of $\kappa$. Thus its corresponding cumulative
distribution function (CDF) can be expressed as
\begin{equation}
C(>\kappa)=\int_{\kappa}P(\kappa^{\prime}) d\kappa^{\prime}.
\end{equation}
where $P(\kappa)$ is normalize to be unit in our analysis. Our
purpose is to extract the PDF of $\kappa$ from noisy $\kappa$ map
and compare it with that of pure convergence field, thus we can
test the method mentioned in last section and apply it to the
analysis of real week lensing data. Using the matrix method, we
reconstruct the PDF of convergence from the pure, noisy $\kappa$
map and pure Gaussian noise smoothed by a Wiener filter.
Fig.(\ref{fig4}) plot the PDFs of the noise-free $\kappa$ field
(solid line), noisy $\kappa$ field (dashed line) and pure Gaussian
noise field (dotted line) respectively. The corresponding error
bars are obtained from averaging over the 40 respective maps. The
empty square dots represent the reconstructed PDF of the $\kappa$
field using the simulated pure $\kappa$ field as the initial
signal field. We see that the reconstructed PDF of $\kappa$ is in
good agreement with that of simulated pure $\kappa$ field, while
there exists slight deviation for smaller negative and larger
positive $\kappa$. Fig.(\ref{fig5}) shows their corresponding
CDFs.

\begin{figure}
\epsscale{1.} \plotone{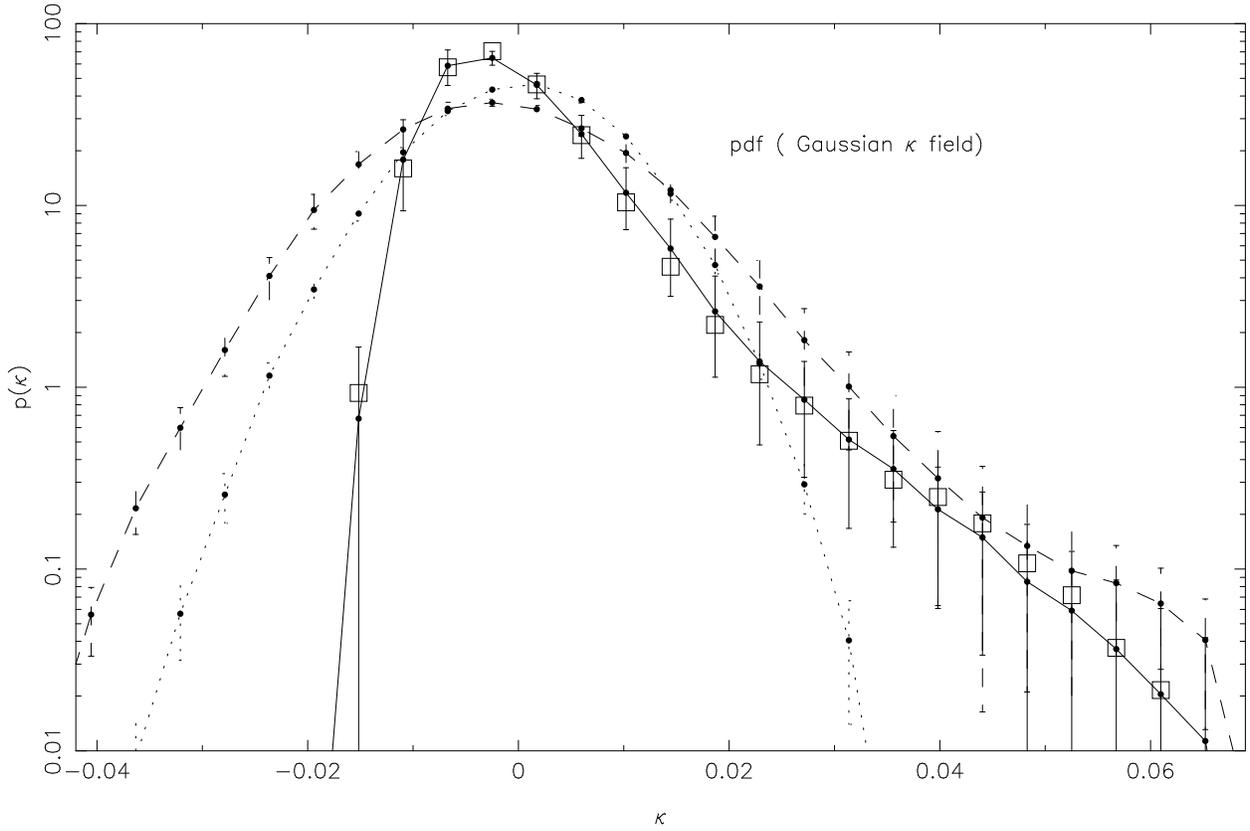} \caption{The same as
Fig.\ref{fig4}, but using a Gaussian field with a mean standard
deviation of 40 simulated pure $\kappa$ maps as an initial signal
field.} \label{fig6}
\end{figure}

\begin{figure}
\epsscale{1.} \plotone{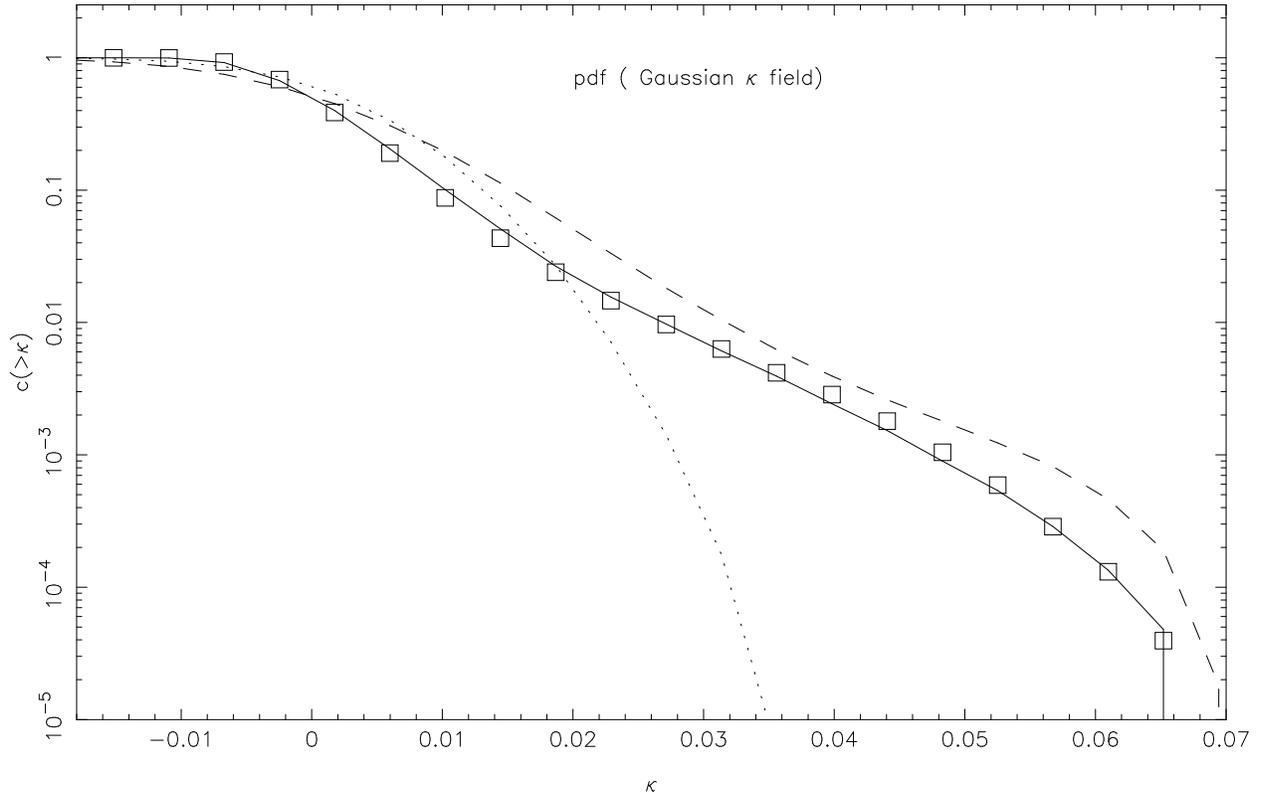} \caption{The same as
Fig.\ref{fig6}, but for the corresponding cumulative distribution
functions (CDF).} \label{fig7}
\end{figure}

\begin{figure}
\epsscale{1.} \plotone{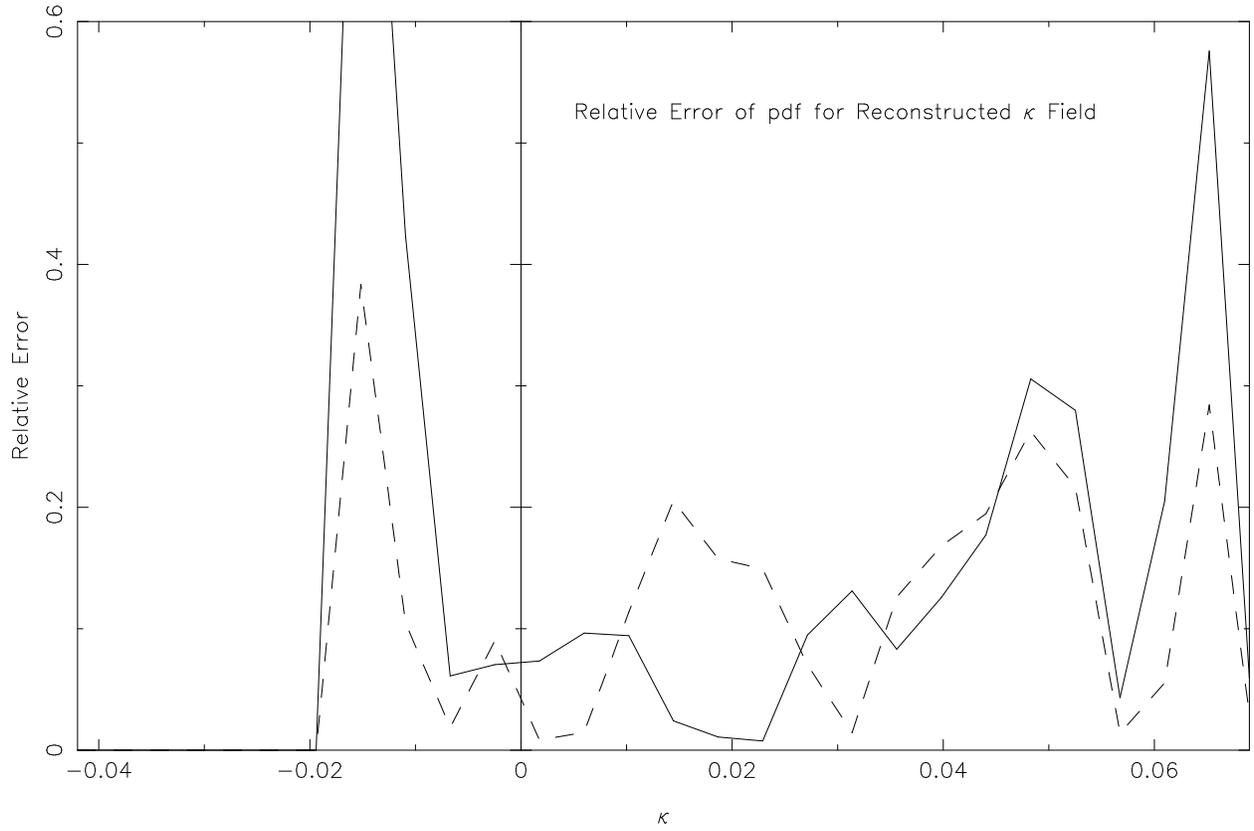} \caption{The relative errors of
PDF for reconstructed $\kappa$ field using an initial signal field
of a simulated pure $\kappa$ field (solid line) and a Gaussian
field (dashed line) respectively.} \label{fig8}
\end{figure}

In the reconstruction process, we used the PDF of the simulated pure
$\kappa$ field as a initial signal field. However, in the real
observation of weak lensing, the PDF of the pure convergence is
unobservable. Thus we employ an approximation that the PDF
$\bar{P}^S(\kappa_j)$ of simulated convergence is replaced by a
Gaussian distribution with mean standard deviation
$\mathbf{\bar{\sigma}}_\kappa$ over 40 simulated $\kappa$ fields,
because we can infer the mean standard deviation of $\kappa$ from
the observational data of weak lensing. Fig.(\ref{fig6}) and
(\ref{fig7}) plot the reconstructed PDFs and their corresponding
CDFs using a Gaussian field with a mean standard deviation of 40
simulated pure $\kappa$ maps as an initial signal field, while
Figs.(\ref{fig8}) shows the relative errors of the PDF for the
reconstructed $\kappa$ field using an initial signal field of a
simulated pure $\kappa$ field (solid line) and a Gaussian field
(dashed line) respectively. Apart from in smallest negative and
largest positive convergence bins, the relative errors for the two
initial PDF distributions are similar and about below 10$\%$.
Therefore, using a Gaussian field PDF with a mean standard deviation
of 40 simulated pure $\kappa$ maps as an initial signal field is a
good approximation in the reconstruction of convergence PDF using
this Wiener reconstruction approach.

\section{Conclusions and Discussions}

In this letter, we ran a high-resolution N-body simulation of a WMAP
cosmology to study the reconstruction of the convergence PDF from
noisy weak lensing data. We added noise due to intrinsic ellipticity
of background faint source galaxies to the simulated $\kappa$ fields
and smoothed it using a Wiener filter.  From the noisy simulated
$\kappa$ field, we make the reconstruction of pdf of convergence by
means of the PDF Wiener method described above. We find the
reconstructed PDF of the convergence are in good agreement with that
of a noise free $\kappa$ map smoothed by a Wiener filter, and its
relative error is below about 10$\%$. We can safely apply this
reconstructed method to the analysis of real observational data for
week lensing such as future CFHT Legacy Survey.

We thank the anonymous referees for their many valuable
suggestions. We also thank Ting-Ting Lu for providing the simulated
lensing data and
Hugh Merz for providing his simulation code. T.J.Zhang would like
to thank Peng-Jie Zhang for useful discussion and Yuan Qiang for
his help on MATLAB, and thank CITA, Xiang-Ping Wu, Xue-Lei Chen,
Bo Qin and Da-Ming Chen for their hospitality during his visits to
the Canadian Institute for Theoretical Astrophysics(CITA),
University of Toronto and the cosmology groups of the National
Astronomical Observatories of P.R.China, respectively.  This work
was partly supported by the National Science Foundation of China
(Grants No.10473002 and 10273003), the 985 Project and the Scientific Research
Foundation for the Returned Overseas Chinese Scholars, State
Education Ministry. The research was also supported in part by
NSERC and computational resources of the CFI funded
infrastructure.

\bibliography{ztjcos-lensbib}
\bibliographystyle{apj}

\appendix

\end{document}